\documentstyle[12pt]{article}
\input{epsf}
\newlength{\extraspace}
\newlength{\extraspaces}
\newcommand{\bea}{\begin{eqnarray}}
\newcommand{\eea}{\end{eqnarray}}
\newcommand{\beq}{\begin{equation}}
\newcommand{\eeq}{\end{equation}}
\newcommand{\brr}{\begin{array}}
\newcommand{\err}{\end{array}}

\newlength{\figsize}
\begin{document}

\thispagestyle{empty}
\begin{flushright}
{\sc OUPT}-01-30-P\\
hep-th/0107163 \\
June 2001 
\end{flushright}
\vspace{.3cm}

\begin{center}
{\Large\sc{ H-BRANES AND CHIRAL STRINGS}}
\\ [12mm]
 {\sc Ian I. Kogan\footnote{i.kogan@physics.ox.ac.uk} and Nuno B.
B. Reis
\footnote{reis@thphys.ox.ac.uk}}\\
\vspace{.3cm}
{Theoretical Physics, Department of Physics \\[2mm]
 University  of Oxford, 1 Keble Road \\[2mm]
 Oxford, OX1 3NP, UK}
\\[12mm]

{\sc Abstract}
\end{center}
\noindent
We add a simple boundary term to the Polyakov action and construct a
new class of D-branes with a single null direction. 
On the string world-sheet the system is described by a single quantized  
left-mode sector of a conformal field theory. By a Wick rotation of
spacetime, we map open strings attached  to  null branes  
into chiral closed strings. We suggest that these so-called  H-branes  
 describe  quantum  horizons - black hole, cosmological (de-Sitter), etc. 
We show how  one can get a space/phase space transmutation
near the horizon and discuss the new features of boundary states which
 become squeezed states.
\\
PACS number(s): 04.20.Dw, 04.50.+h, 11.25.Mj, 98.80.Hw \\

\newpage
\renewcommand{\footnotesize}{\small}


%
%

\section{Introduction}

 It was shown in the  early seventies by Christodoulou \cite{christo},
Penrose and Floyd \cite{penrose} and Hawking \cite{hawking71}  
that the horizon area of a black hole cannot decrease. 
 In his seminal paper \cite{bekenstein1} Bekenstein used  this
  to identify horizon area with a black hole  entropy which can be
 defined as the
 measure of information  inaccessible to an exterior observer. It was
 also shown  that the minimum  increase of the black hole area  has to be
proportional to the square of Planck length. 
Later, Bardeen, Carter and Hawking found the 
four laws of black hole mechanics to be analogous to the usual four
laws of thermodynamics \cite{hawking1}.
 This  was finally realized to be the correct interpretation when
Hawking found that black holes do in fact radiate energy 
and behave as  hot objects with temperature 
proportional to the surface gravity
\cite{hawking2}. 
Thus the mechanical laws of black holes are thermodynamical laws and
one can 
identify the entropy of any black hole with 
the  one quarter of the it's horizon area (in Planck units) 
\bea
S_{BH} = \frac{1}{4} A 
\eea
The next important step  was the suggestion of  Bekenstein 
   that the  area $A$ (and the mass) must be
also quantized in Planck units \cite{bekenstein2}. 
This conjecture was based on earlier observations
\cite{christo, chrsitoruffini} that the horizon area of a
non-extremal black hole behaves as an adiabatic invariant. 
The   minimal increase of
 a black hole area   is an insight  that  quantum
horizon  plays the 
role of a phase space and  one can 
 form   independent patches of equal Planck size areas, bearing 
in mind that each patch can locate one degree of freedom as in a usual
 phase space patchwork - what is called now Holographic principle
 \cite{thooft},\cite{susskind6}. 

The idea of a discrete spectrum proposed by Bekenstein, was later
discussed  independently by Mukhanov \cite{muk} and one of the authors 
 (IIK) \cite{kogan1} \footnote{Unfortunately at that time both VM and
IIK were  unaware about Bekenstein pioneering paper
\cite{bekenstein2}.}.  Arguments used in \cite{muk} were based on
entropy and further developments are  discussed in \cite{bekmuk}. 
In paper  \cite{kogan1} (see also \cite{kogan2}) a stringy approach to 
quantization  was suggested. It was based on a consistency
 of a {\it chiral} sector of a closed string moving in a Euclidean black hole
 background  where the time coordinate is periodic.
The fact that it leads to the same discrete spectrum as Bekenstein's 
 was an indication that  chiral sectors should be relevant to the
statistical  counting of black hole entropy. Let us  note that in
 the  case of extremal Kerr-Newman black holes the quantization of
mass  follows from the quantization of electric and magnetic charges and 
angular momentum \cite{mazur}.
 In recent  years the discrete spectrum  of quantum black holes
 was discussed in numerous papers,
see for example the recent papers \cite{bhrecent} and references therein. 
 Let us also note that quantization of an  area operator 
 has been  extensively  discussed recently  in Ashtekar's 
approach to  quantum
gravity - see \cite{ars} and references therein.

The same paradigm about single chiral sector near  horizon  was later
suggested by Carlip  \cite{carlip1}  first for  BTZ (2+1) black hole
\cite{banados1,banados2} and  for general black holes in \cite{carlip3}.
One starts to study the quantum version of these black holes since it
is well known that (2+1)-dimensional gravity can be written as a
 topological  Chern-Simons theory \cite{townsend,witten,carlip2} 
 which  justify   the Holographic Principle
\cite{thooft, susskind6} in 2+1 dimensions and one  can say
 that  entropy of 
BTZ-black holes  is determined by some quantum states  which 
are restricted to live on a (1+1)-dimensional surface. 
In (2+1)-dimensional gravity, the presence of a boundary is in 
fact necessary since the bulk by itself does not contain enough
gravitational 
degrees of freedom to account for black hole entropy \cite{carlip2}. 
There are two natural candidates for this boundary surface, the
 horizon itself and spatial infinity. 

Carlip considers the BTZ black hole horizon to be a true boundary where
usual invariant diffeomorphisms are broken, 
 turning "would be pure gauge of freedom" into  physical ones
\cite{carlip1}. 
In the Chern-Simons  description of (2+1)-gravity one gets
 chiral sector of a WZW theory of level $k$ which lives on
 a boundary.  The level $k=\frac{l\sqrt 2}{8G}$ is  
 related to the
cosmological constant 
$\Lambda=-1/l^{2}$ of the AdS spacetime. In  the quasi-classical
 approximation  a  conformal field theory on a boundary  has the central 
charge $c=6$. The Virasoro operator $L_{0}$ takes the form \cite{carlip1},

\beq
L_{0} \sim N - \left( \frac{\rho_{+}}{4G} \right)^{2},\ ~~~~~~~~~ 
\rho_{+}^{2}= 4Gl \sqrt{Ml^{2}-J^{2}} + 4GMl^{2}
\eeq
where $\rho_{+}$ is the value of the parameter in front of the usual
$\phi$-periodic 
coordinate on the BTZ metric in Lorenz signature. 
Using Cardy formula \cite{cardy1} for number of states  at
 a given level $N$ in  a CFT with
 a central charge $c$ 
\beq
n(N) \sim exp \left[ 2\pi \sqrt{c\frac{N}{6}}\right]
\eeq
one can get entropy $ S = ln \,\,n(N)$.
Black hole states 
are  determined by on-shell condition  $L_{0}=0$ which gives entropy
\beq
S =  \pi \sqrt{c\frac{2N}{3}}  = \frac{2 \pi \rho_{+}}{4G} 
\eeq
in agreement with the Bekenstein-Hawking BTZ-black hole entropy. As
was shown in \cite{kogan2}  $L_{0}=0$ condition also leads to
$\rho_{+}$ and mass quantization. 

Strominger has given  
another approach to count the number of states of the BTZ black holes
by considering the boundary as a (t,$\phi$) 
cylinder at spatial infinity \cite{strominger3}. 
The CFT at this boundary  is  a  Liouville with central charge
$c=\frac{3l}{2G}$ \cite{coussaert}.
 Considering the ground state of it as the AdS spacetime, Strominger 
has made the following identifications of Virasoro operators with the
mass 
and angular momentum of the black hole 
\bea
L_{0}+ \bar{L_{0}} = lM, ~~~~~~~~~~
L_{0}- \bar{L_{0}} = J
\eea

The correct value of the entropy is given by considering the
contributions from both sectors of the theory:
\beq
S=\sqrt{c \frac{L_{0}}{6}}+\sqrt{c \frac{ \bar{L}_{0}}{6}}= \frac{2 \pi \rho_{+}}{4G}
\eeq

We see that there are at least two different approaches to count the
number of states (see also \cite{carlip3,ash,sol,carlip4}).
 The  BTZ black hole example teaches us  the following lesson:
the contribution  of  states to the  entropy comes
from just one Virasoro algebra ( a chiral sector)   or from two
Virasoro algebras (non-chiral sector)  depending on how we choose the
boundary - at the horizon or at infinity.

Let us now turn to another  aspect of strings in black hole
background. 
There is an apparent inconsistencies of what two distinct observers measure in the
presence of a black hole,
 where one is freely falling to the black hole and the other is fixed
outside it \cite{susskind3}. 

First, the different rates on the clocks between these two observers
from the red-shift phenomena, 
has consequences on the different sizes that they measure in a string
falling to the horizon \cite{susskind1}. 
That is because the size of the string depends on how many excitation
modes one is able to see  and such number depends on the resolution
time 
$\epsilon$ - in Planck units - of the observation. 
As the string is examined with better and better time resolution it
appears 
to slowly grow as shown by the relation,
\beq
R^{2}_{string}\sim \alpha ' \log\frac{1}{\epsilon}
\eeq
 For the freely falling observer the transverse size of the string is
always of the 
order the Planck scale since his resolution in time will not be smaller
than 
the Planck time. However, the outside  observer is able to see
more excitation modes 
as the string approaches the horizon because of the red-shift factor on
his clock. String starts to spread over the space and consequently 
 the external observer sees a stretched horizon  formed by the strings
  frozen near horizon.  
The stretched horizon is a hot place with a temperature near Hagedorn
limit. Black hole - string complementarity based on this picture was
discussed in  \cite{susskind1,damour1,damour2}.
 
Second, a closed string on the horizon seen by the freely falling
observer can be seen as an open string by the outside fixed one, since
she/he is not able to see what is inside the black hole. Susskind
conjectured that the classical Bekenstein-Hawking black hole entropy
arises  from configurations of these open strings with the ends frozen
at the horizon. 
Quantum corrections of the classical entropy are expected to be finite
at all orders in string perturbation theory \cite{uglum2}.

Third, the different behaviors of the string size growth indicate a
possible non-commutativity 
of the light-cone coordinates and to the resolution of the information
loss paradox \cite{susskind7}. Strings carry information on their
oscillators and  one can argue that information has spread along the
stretched horizon with non-commutative  light-cone coordinates,

\beq
\left[X^{-},X^{+}\right]=\alpha'
\eeq
From  this relation we can  see  that by taking 
smaller and smaller time intervals  information starts to spread over
the space, as shown in Figure 1.

\begin{figure}
\hspace{35mm}
\epsfxsize=63mm
\epsfysize=77mm
\epsffile{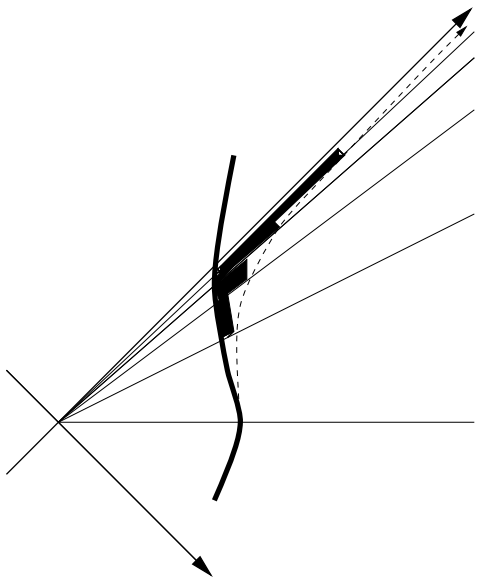}
\caption{Information spreading along the stretched horizon. 
All squares have equal area by the Heisenberg inequality}
\label{figure:fig1}
\end{figure}

String theory was able to give us a partial answer of how to count the
states that give the one-quarter-of-the-area 
law by using D-brane technology on extremum black holes, with a near
AdS-geometry horizon \cite{strominger1,maldacena}.
 The counting of states is not done by a quantum gravity describing
the geometry of near horizon but by a dual theory 
of supergravity in the UV limit on AdS spacetime.
 That is the IR limit of a gauge theory  that lives on D-branes
\cite{maldacena2}. 
In such a limit, the black hole has become smaller then the string scale 
and so there is no concept of a horizon in this dual description
since 
spacetime geometry is quite fuzzy at IR scales. 
Thus it is difficult to understand how one can explain the 
universality of the Bekenstein-Hawking entropy law \cite{strominger2}.
\begin{figure}
\hspace{70mm}
\epsfxsize=40.5mm
\epsfysize=63mm
\epsffile{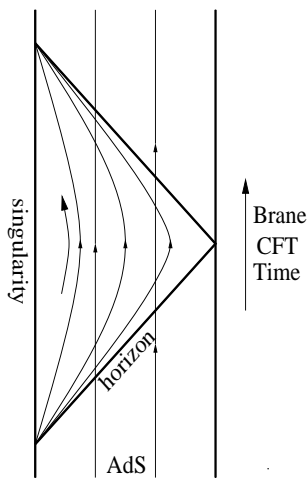}
\caption{Extremum Black Hole geometry immersed in a commutative AdS spacetime.}
\label{figure:fig2}
\end{figure}
  Such dual description is also problematic in terms of black hole
unitary evolution. Figure 2 shows the embedding of the whole extremum 
black hole geometry
in a AdS spacetime \cite{strominger2}.
 Curved rows represent the black hole time isometrics and the straight
lines are time evolution in 
 AdS spacetime with  D-branes inside. 
From the AdS/CFT dual picture, 
straight lines describe time evolution of a unitary conformal field
theory. 
However it is not obvious why black hole evolution should be
unitary 
from the dual picture point of view. Initial configuration evolves into the
black 
hole and forgets  about the information that is in the rest of AdS
spacetime. 
Moreover, why information cannot flow outside the black hole and go 
through the rest of AdS spacetime and continue to propagate freely 
until it reaches infinity? Conceptually the  dual picture seems 
not to  give a {\it universal} description of a Bekenstein-Hawking entropy
  and we are still far from  understanding it 
microscopic origin .

So far we mostly addressed an issue of black hole horizons.
Another important issue is a stringy description of cosmological
horizons. In the  case of de-Sitter  space this problem  recently attracted
a lot of attention - see \cite{desitter} and references therein. Again 
we have to find where are the degrees of freedom live which
contribute to the entropy of cosmological event horizon in de-Sitter
space  \cite{GibbonsHawking}.

From the previous discussion we learnt one important thing.
It seems that when we try to get a stringy description of a quantum 
horizon we  either have to deal with open strings attached to this
horizon  or with a chiral sector of closed strings. The only common
feature of these two systems is that both have {\it one} Virasoro
algebra. Thinking about  an open string description one can suggest a 
D-brane \cite{polchinski1} stretched along a horizon. However it will
not work.  As we shall demonstrate later it is impossible to   get a 
 usual separation between Neumann and Dirichlet directions. 
 More importantly  if one has open strings on a D-brane they will be open whatever
analytical continuation we are going to undertake.  But we need
something which gives us open  strings in Lorentzian signature and 
 at the same time chiral closed strings after continuation into
the Euclidean domain. One should  remember that the horizon itself shrinks to
a point when one passes  Lorentzian signature to a Euclidean one.

In this paper we suggest a new type of branes - {\bf H-branes}, which
 have one single null direction in a space with a Lorentzian signature
 and which naturally support chiral closed strings after analytical
continuation. We shall focus our attention on the  properties of these branes
 in  flat space (leaving actual applications to black holes and
cosmological models for future publications).

The paper is organized as follows. 
In the next section  we review the concept of a bosonic string moving 
in spacetime and  add an extra boundary term to the Polyakov action
 which will lead to light-like  world-sheet boundaries.
In Section 3 we introduce H-branes and 
obtain  the equations of motion  that open strings must obey in
Minkowski spacetime in a presence of  H-brane. 
In Section 4  we perform a Wick rotation in both  spacetime and world-sheet
to obtain chiral closed strings from open strings ending on a
H-brane.
In section 5  an analogy between our model and a Dissipative
Quantum Mechanics is discussed as well as  Ishibashi states associated
 with  our new boundary conditions.  
We also show that boundary state for H-brane is not  a coherent but
a squeezed state. In  conclusion we discuss 
open questions to be addressed in the future.

\section{Open Strings with  light-like world-sheet boundaries}

  We  start by asking the following question: is it possible to
have an open string which has a   Neumann condition for let say $
X^{+}$ and  a Dirichlet condition for $X^{-}$, where  $X^{\pm}$ are
 target space light-cone coordinates ?

To answer this question  let's recall how one gets boundary conditions for open
strings (for more details see \cite{polchinski3}). The starting point 
is  the variation of the  Polyakov action 
\beq
S_{P}=-\frac{1}{4\pi\alpha'}\int d\tau d\sigma 
(-\gamma)^{1/2} \gamma^{ab} \partial_{a}X^{\mu} \partial_{b}X_{\mu}
\eeq
which gives   equations of
motion  from the bulk term  plus  boundary conditions from the boundary term
\beq
\delta S_{P}=\frac{1}{4\pi\alpha'}\int d\tau
d\sigma  \delta X^{\mu} \partial_{a}\left( (-\gamma)^{1/2} \gamma^{ab}
\partial_{b}X_{\mu} \right) -
\frac{1}{4\pi\alpha'}\int_{\partial \Sigma} ds \left( \delta X^{\mu}
t^{a} \partial_{a}X_{\mu} \right)
\eeq
Boundary term is given by the  second  integral which is  taken over 
the world-sheet
boundary $\partial \Sigma$ and  $t^{a}$ is a unit tangent vector.

Let us  choose  space-time light-cone coordinates:
\beq
X^{\pm}=\frac{1}{\sqrt2}(X \pm T)
\eeq
and choosing the usual world-sheet metric in conformal gauge with
world-sheet  boundaries as  space-like lines. The  the boundary conditions are,
\bea
&&\delta X^{+}\partial_{\sigma}X^{-}=0\nonumber\\
&&\delta X^{-}\partial_{\sigma}X^{+}=0\nonumber\\
&&\delta X^{i}\partial_{\sigma}X^{i}=0
\label{boundaryconditions}
\eea
where $i=2,3,...,25$ is an index of the transverse coordinates (in
this paper we only consider bosonic strings).

From the first line of (\ref{boundaryconditions}), 
we see that if we choose the Dirichlet condition for $X^{+}$, then  $
\partial_{\sigma}X^{+}\neq 0$. The  from the second line we are forced
to  choose also a Dirichlet condition for $X^{-}$, since $X^{+}$ don't
obey a Neumann condition. 
By the same argument, if we  choose the Neumann condition for $X^{-}$, 
then we will end also with a Neumann condition for $X^{+}$.
 The reason is very simple - to have either Dirichlet or   Neumann
condition for  let say  $X^{+}$ we must have it for both $X$ and $T$,
but then we shall get the same condition for  $X^{-}$.
 On  transverse coordinates there are no such  restrictions
since boundary conditions 
don't have a cross structure as in $+,-$ case. 
We can choose for each transverse coordinates, either Dirichlet or
Neumann conditions. 
Thus, even if we know that the endpoints of an open string move at
speed of light, 
the previous argument show us that is not possible to attach both ends
on a single null direction. 
It seems  a D-brane with a single null direction is impossible.

Is it really so ?

\subsection{World-sheet with light-like boundaries}

We are going to introduce a new boundary term  which
 will correspond to a new type of world-sheet boundaries for open
strings. To be more precise we shall get a light-like boundary which
 is the key element to get a brane with a single null direction.
Let us start from postulating an extra boundary term

\beq
S_{B}=\frac{1}{8\pi\alpha'}\int_{\partial \Sigma} ds 
\left( X^{+} t^{a} \partial_{a}X^{-}- X^{-} t^{a} \partial_{a}X^{+} \right)
\eeq
where the integral is taken over the world-sheet boundary  and 
 $t^{a}$ is a unit tangent vector.
 This  term is Weyl and diffeomorphism  invariant  
 so  one can add  it to the Polyakov action.
 Let us  choose a coordinate system 
such that the variable $\tau$ parameterizes the world-sheet boundary
 $\partial\Sigma$. The infinitesimal proper distance $ds$ is defined 
by $ds=(-\gamma)^{1/4}d\tau$.
Since under an infinitesimal Weyl transformation 
the world-sheet metric is changed as $
\delta_{W} \gamma_{ab}=2\gamma_{ab}\delta\omega
$
we see that ds is changed in the following manner
$\delta_{W}ds=ds\delta\omega$
where we have kept the coordinate $\tau$ unchanged. 
The tangent vector $t^{a}$ on the extra boundary term is changed by a 
Weyl transformation since we have to keep it as a unit vector 
\beq
\delta_{W} \left( \frac{t^{a}}{|t|} \right) =-\frac{t^{a}}{|t|}\delta\omega.
\eeq
The partial derivatives are not changed by the Weyl transformation
since  they only depends on the world-sheet coordinates $\sigma^{a}$:

\beq
\delta_{W}\frac{\partial}{\partial\sigma^{a}}=0
\eeq
The proper distance ds and the unit tangent vector $t^{a}$ are changed
by 
the same magnitude but with different signs so the total contribution
to the infinitesimal Weyl transformation on the extra boundary term is null

\beq
\delta_{W}S_{B}=0
\eeq
In a similar way one can show that  this  boundary term  is invariant
under diffeomorphisms. However, the  boundary term is no longer
spacetime Poincare invariant as it depends on the explicit null
spacetime coordinates. As we  shall demonstrate, this arises from 
 the presence of a (null) brane. The total action is 

\beq
S_{P+B}=-\frac{1}{4\pi\alpha'}\int d\tau d\sigma (-\gamma)^{1/2} 
\gamma^{ab} \partial_{a}X^{\mu} \partial_{b}X_{\mu}-
\frac{\beta}{4\pi\alpha'}\int_{\partial \Sigma} ds X^{+} t^{a} 
\partial_{a}X^{-}
\eeq
where  we integrated the boundary term by parts and introduced a free
parameter  $\beta$. Both the  world-sheet and the target space 
 have  Lorenz signature.

 The action  has a local
Weyl$\otimes$Diff  symmetry  and and one can impose a three gauge
conditions on our world-sheet metric. 
We will work in light-cone gauge for the world-sheet time coordinate 
and impose two more conditions defined as follows \cite{polchinski3}:
\bea
X^{+}=\tau, \,\,\,\,\,\,
\partial_{\sigma} \gamma_{\sigma \sigma}=0,  \,\,\,\,\,\,
&&\gamma=-1. 
\eea
The procedure to choose the above conditions is  discussed in details
in \S 1.3 of \cite{polchinski3} where one defines the world-sheet 
coordinate $\sigma$ by constructing it from one boundary $\sigma=0$ to 
another one $\sigma=\pi$ using an invariant length 
$dl=\gamma_{\sigma \sigma}(-\gamma)^{-1/2}d\sigma$.
Under the above conditions and by splitting $X^{-}$ into  mean
value 
\bea
{\tilde X}^{-}(\tau)=\frac{1}{\ell}\int_{0}^{\ell}d\sigma
X^{-}(\tau,\sigma), 
\eea
and $Y^{-}(\tau,\sigma)=X^{-}(\tau,\sigma)-{\tilde X}^{-}(\tau)$
 one can see that $Y^{-}$ acts as a Lagrange multiplier 
constraining $\partial_{\sigma}\gamma_{\tau \sigma}$ to vanish. 
The presence of the new  boundary term in the action can be seen as a
deformation of the original  Lagrangian  by boundary vertex operators
\beq
L_{P+B}=L_{P}+\frac{\beta}{4\pi\alpha'}\left(X^{+}\partial_{\tau}X^{-}
\mid_{\sigma=0}+X^{+}\partial_{\tau}X^{-}\mid_{\sigma=\pi} \right)
\eeq
but this deformation  does  not change the abovementioned  constraint
 $\partial_{\sigma}\gamma_{\tau \sigma}=0$. 
But open string boundary conditions are modified:
\bea
\gamma_{\tau \sigma}\partial_{\tau}X^{+}-
\gamma_{\tau \tau}\partial_{\sigma}X^{+}-\beta\partial_{\tau}X^{+}=0\nonumber\\
\gamma_{\tau \sigma}\partial_{\tau}X^{-}-
\gamma_{\tau \tau}\partial_{\sigma}X^{-}+\beta\partial_{\tau}X^{-}=0\nonumber\\
\gamma_{\tau \sigma}\partial_{\tau}X^{i}-
\gamma_{\tau \tau}\partial_{\sigma}X^{i}=0
\eea
where we have used $ds=d\tau$. The different signs between the first
two lines comes from integration by parts when we take the variation
of the  boundary term. 
Using  the light-cone gauge choice $X^{+}=\tau$ one gets 
\beq 
\gamma_{\tau \sigma}=\beta\ ~~ at\ ~~ \sigma=0,\pi.
\eeq
and since $\partial_{\sigma}\gamma_{\tau \sigma}=0$ we have 
$\gamma_{\tau \sigma}=\beta$ everywhere. By the condition $\gamma=-1$, 
we must have at each point that 
$\gamma_{\tau \tau}\gamma_{\sigma \sigma}=\beta^{2}-1$. 
When $\beta=0$ we have the usual world-sheet metric with time-like
boundaries for open strings. However, for $\beta=1$ we have the
condition
\beq
\gamma_{\tau \tau}\gamma_{\sigma \sigma}=0
\eeq
 and the metric takes either the form 
\beq
ds^{2}=\gamma_{\tau \tau}d\tau^{2} + 2 d \tau d \sigma
\eeq
for the case $\gamma_{\sigma \sigma}=0$ or
\beq
ds^{2}=\gamma_{\sigma \sigma}d\sigma^{2} + 2 d \tau d \sigma
\eeq
for  $\gamma_{\tau \tau}=0$. 
One can   see that when  $\beta=1$ 
 there are  two type of boundaries (let us remind that along the
boundary  $d\sigma= 0$ )
\bea
&&ds^{2}=\gamma_{\tau \tau}d\tau^{2}\ ~~~~~~~~ for\ ~ 
\gamma_{\sigma \sigma}=0\ ~~ or, \nonumber\\
&&ds^{2}=0\ ~~~~~~~~~~~~~~~~ for\ ~ \gamma_{\tau \tau}=0
\eea
In the first case, boundaries are still time-like 
but  in the second case we have  a  
 totally new  type of light-like boundaries with the metric
\beq
\gamma_{ab}=
\left[
\brr{cc}          
0&1\\
1&\gamma_{\sigma \sigma}(\tau)
\err
\right]
\eeq
In the rest of the paper we shall work with this metric 
 and with  world-sheets with 
{\it light-like  boundaries}.

Let us  find the equations of motion for transverse coordinates.
The Lagrangian is reduced to
\bea
&&L=-\frac{\ell}{2\pi\alpha'}\gamma_{\sigma \sigma} \partial_{\tau} 
{\tilde X}^{-}(\tau)\nonumber\\
&&\ ~~~ +\frac{1}{4\pi\alpha'}\int_{0}^{\ell}d\sigma 
\left[ \gamma_{\sigma \sigma}\partial_{\tau}X^{i}\partial_
{\tau}X^{i}+2(\partial_{\sigma}Y^{-}-\partial_{\tau}X^{i}
\partial_{\sigma}X^{i}) \right]\nonumber\\
&&\ ~~~
\eea
and the conjugate  momenta are: 
\bea
&&p_{-}=-p^{+}=\frac{\delta L}{\delta(\partial_{\tau}{\tilde X^{-}})}=
-\frac{\ell}{2\pi\alpha'}\gamma_{\sigma \sigma}(\tau)\nonumber\\
&&\Pi^{i}=\frac{\delta
L}{\delta(\partial_{\tau}X^{i})}=\frac{p^{+}}{\ell}
\partial_{\tau}X^{i}-\frac{1}{2\pi\alpha'}\partial_{\sigma}X^{i}
\eea
The Hamiltonian is the following:
\bea
&&H=p_{-}\partial_{\tau}{\tilde X^{-}}+\int_{0}^{\ell}d\sigma\Pi_{i}
\partial_{\tau}X^{i}-L\nonumber\\
&&~~~=\frac{\ell}{2\pi\alpha'p^{+}}\int_{0}^{\ell}d\sigma 
\left[\pi\alpha' \Pi^{i}\Pi^{i}+\Pi^{i}\partial_{\sigma}X^{i}+
\frac{1}{4\pi\alpha'}\partial_{\sigma}X^{i}\partial_{\sigma}X^{i}
\right]
\nonumber\\
&&~~~-\frac{1}{2\pi\alpha'}\int_{0}^{\ell}d\sigma 
\partial_{\sigma}Y^{-}
\eea
from where we take the equations of motion
\bea
&&\partial_{\tau}p^{+}=\frac{\delta H}{\delta x^{-}} = 0\nonumber\\
&&\partial_{\tau}\Pi^{i}=-\frac{\delta H}{\delta X^{i}}=
\frac{\ell}{2\pi\alpha'p^{+}}\left[\partial_{\sigma}\Pi^{i}+
\frac{1}{2\pi\alpha'}\partial_{\sigma}^{2}X^{i}\right]
\eea
with $p^{+}$ a conserved quantity. 
From the second line we have a wave equation 
\beq
\partial_{\sigma}\partial_{\tau}X^{i}=\frac{\pi\alpha' 
p^{+}}{\ell}\partial_{\tau}^{2}X^{i}
\eeq
It is  convenient to take $p^{+}=0$ 
($\gamma_{\sigma \sigma} = 0$) and consider the 
$\tau$ and $\sigma$ coordinates as light-like 
\bea
\tau=x^{+}, \,\,\,\,\,\,\,  \sigma=x^{-} 
\eea
so one  can  write the wave equation in a  usual form. 
The world-sheet metric takes now the form $ds^{2}=dx^{-}dx^{+}$ 
with boundaries at $x^{-}=0$ and $x^{-}=\pi$. 
The world-sheet of such configuration is drawn in Figure 3. 
\begin{figure}
\hspace{60mm}
\epsfxsize=55mm
\epsfysize=55mm
\epsffile{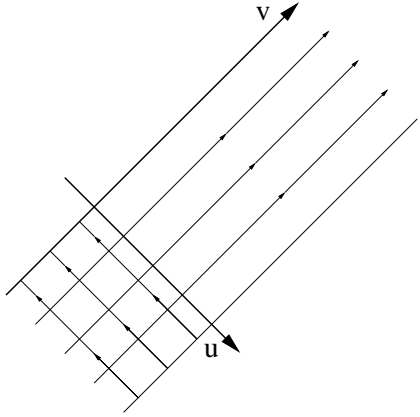}
\caption{String modes propagating on world-sheet, 
where the boundaries are at $x^{-}=0$ and $x^{-}=\pi$}
\label{figure:fig3}
\end{figure} 
In this limit  the Lagrangian takes the form,
\bea
L=\frac{1}{2\pi\alpha'}\int_{0}^{\ell}d x^{-} \left[\partial_{-}Y^{-}-
\partial_{+}X^{i}\partial_{-}X^{i}\right]
\eea
where the partial derivatives refers to the world-sheet 
coordinates $x^{-}$ and $x^{+}$. Since the momentum $\Pi^{i}$ is
\beq
\Pi^{i}=-\frac{1}{2\pi\alpha'}\partial_{-}X^{i}
\eeq
we can see that the Hamiltonian takes unusual form
\beq
H=-\frac{1}{2\pi\alpha'}\int_{0}^{\ell}dx^{-} 
\partial_{-}Y^{-}
\eeq
Let us note that in the 
limit $p^{+}\rightarrow 0$  all transverse oscillators are frozen.

\section{H-Branes}

Let us study an open string on this new world-sheet with light-like
 boundaries induced by the special boundary term we have introduced.
Taking the variation of the  Polyakov action 
\beq
\delta S_{P}=\frac{1}{4\pi\alpha'}\int_{\Sigma} dx^{-} dx^{+} \delta
X^{\mu}
 \partial_{-}\partial_{+}X_{\mu}-\frac{1}{4\pi\alpha'}\int_{\partial
\Sigma}dx^{+} \delta X^{\mu}\partial_{+}X_{\mu}
\eeq
and  adding the variation of the  boundary term
\bea
\delta S_{b}=-\frac{1}{4\pi\alpha'}\int_{\delta \Sigma} dx^{+} 
\left[\delta X^{+} \partial_{+}X^{-}+X^{+}\partial_{+}\delta
X^{-}\right]
\nonumber\\
=-\frac{1}{4\pi\alpha'}\int_{\delta \Sigma} dx^{+} 
\left[\delta X^{+} \partial_{+}X^{-}-\delta X^{-}\partial_{+}X^{+}\right]
\eea
one can see that the  total contribution  involving target space
light-cone coordinates is
\bea
&&+\delta X^{+} \partial_{+}X^{-}+\delta X^{-} 
\partial_{+}X^{+} \ ~~~ from\ Polyakov\ action\nonumber\\
&&+\delta X^{+} \partial_{+}X^{-}-\delta X^{-} 
\partial_{+}X^{+} \ ~~~ from\ extra\ boundary\ term\nonumber\\
&&=2\delta X^{+} \partial_{+}X^{-}
\eea
Thus we end with just one boundary condition on the light-cone coordinates
\beq
\delta X^{+} \partial_{+}X^{-} \mid_{\partial \Sigma}=0
\label{newcondition}
\eeq
 At the same time nothing changed for the 
transverse coordinates $X_i$,  $i=2,3,...,25$
\beq
\delta X^{i} \partial_{+}X^{i} \mid_{\partial \Sigma}=0 
\eeq

From  the boundary condition (\ref{newcondition}) we
are free to choose either $\delta X^{+}=0$ or $\partial_{+}X^{-}=0$
 - contrary to what we have had for open strings  with time-like boundaries.
 Thus we can have the endpoints of the
open string fixed in a single null $X^{+}$ direction with no boundary
condition for $X^{-}$ or vice-versa. This defines a   new type of  branes
   which we shall call {\bf H-brane}. \footnote{ A second  choice
was N-brane, where N stands for null, but we prefer to call it
H-brane stressing the natural relation to horizons.  Let us note that
nevertheless there is some information about null structure in the
letter H also - if you read it in Russian ( Russian H $=$ Latin N).}

Equations of motion for $X^{\pm}$ are
\beq
\partial_{-}\partial_{+}X^{\pm}=0
\eeq
have  general solutions 
\bea
&&X^{-}(x^{-},x^{+})=X^{-}_{L}(x^{-})+X^{-}_{R}(x^{+})\nonumber\\
&&X^{+}(x^{-},x^{+})=X^{+}_{L}(x^{-})+X^{+}_{R}(x^{+})
\eea
The second term of each light-cone
coordinate is not restricted to discrete spectrum (we do not insist on 
periodicity in $x^{+}$)
\bea
&&X^{-}_{R}(x^{+})=X^{-}_{R}(0) + p^{-}_{R}x^{+} + (4\pi\alpha')^{1/2} 
\int \left[ \frac{a^{-}_{n}}{n}\cos(nx^{+})+
\frac{b^{-}_{n}}{n}\sin(nx^{+}) 
\right]dn \nonumber\\
&&X^{+}_{R}(x^{+})=X^{+}_{R}(0) + p^{+}_{R}x^{+} + (4\pi\alpha')^{1/2} 
\int \left[ \frac{a^{+}_{n}}{n}\cos(nx^{+})+
\frac{b^{+}_{n}}{n}\sin(nx^{+}) 
\right]dn\
\eea
Let us take  the Dirichlet boundary condition  for $X^{-}$
\bea
\partial_{+}X^{-}=0
\eea
This leads to constraints  for the oscillators as well as for the
 momentum $p^{-}$
\bea
\int a^{-}_{n}\sin(nx^{+})dn -\int b^{-}_{n}\cos(nx^{+})dn=0\nonumber\\
p^{-}_{R}=0
\eea
at $x^{-}=0$ and $x^{-}=\pi$. Since the condition must be valid at 
any light-like time $x^{+}$, the oscillators $a^{-}_{n}$ and
$b^{-}_{n}$ 
must all vanishes, as well as the $p^{-}$ momentum. 
Inserting the values of $X^{\mu}_{R}(0)$ into $X^{\mu}_{L}(x^{-})$, 
we have the solution
\bea
&&X^{-}(x^{-},x^{+})=X^{-}_{L}(x^{-}), \,\,\,\,\,\,\,
X^{i}(x^{-},x^{+})=X^{i}_{L}(x^{-}) \\ 
&&X^{+}(x^{-},x^{+})=X^{+}_{L}(x^{-})+p^{+}_{L}x^{+} + 
(4\pi\alpha')^{1/2}\int \left[ \frac{a^{+}_{n}}{n}\cos(nx^{+})+
\frac{b^{+}_{n}}{n}\sin(nx^{+}) \right]dn\nonumber
\eea 
The light-cone coordinates at the endpoints of 
the open string have some fixed values for $x^{-}=0$ 
and $x^{-}=\pi$. Note that the Dirichlet boundary condition 
$\partial_{+}X^{-}\mid_{\partial \Sigma}=0$ 
restricts the endpoints of an open string to be fixed at 
the $X^{-}$ light-cone coordinate
\beq
\partial_{+}X^{-}
\mid_{\partial \Sigma}=0 \Rightarrow \delta X^{-}\mid_{\partial \Sigma}=0
\eeq
The same is true for the transverse coordinates, 
as we have the same equations of motion and boundary conditions.
 We just don't have any boundary condition for $X^{+}$. 
Thus, we end up with a string that has it endpoints frozen 
 along the null $X^{-}$ direction as well as in
 all the transverse directions, 
but is allowed to move along the $X^{+}$ direction. 
Of course one can swap  $X^{-}$ and  $X^{+}$ in this construction.

The solution $X^{-}(x^{-},x^{+})=X^{-}_{L}(x^{-})$ 
admits open strings attached to a single H-brane as well as an open
string with endpoints attached to  two different H-branes and there
are two cases:
\bea
&&X^{-}_{1}(x^{-}=0)=X^{-}_{1}(x^{-}=\pi)\nonumber\\
&&X^{-}_{2}(x^{-}=0)=X^{-}_{2}(x^{-}=\pi)+D^{-}
\eea
with $D^{-}$ a proper distance along the null direction $X^{-}$
between the two H-branes, as Figure 4 shows.
\begin{figure}
\hspace{60mm}
\epsfxsize=55mm
\epsfysize=55mm
\epsffile{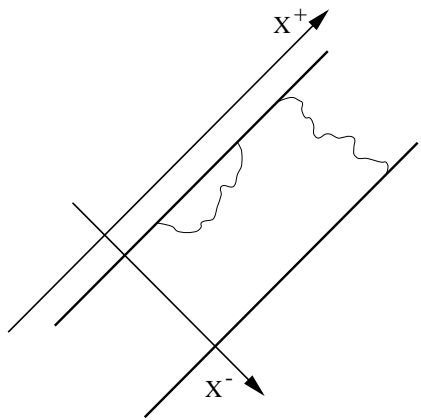}
\caption{Open strings attached to H-branes in Minkowski spacetime.}
\label{figure:fig4}
\end{figure}
In terms of world-sheet picture, oscillating modes evolve parallel to 
the light-like boundaries without any reflection on it. Because
The energy momentum tensor $T_{ab}$  has only one non-zero component
\bea
T_{--}=-\partial_{-}X^{\mu}\partial_{-}X_{\mu}, \,\,\,\,\,
T_{-+}=0, \,\,\,\,\,
T_{++}=0
\eea 
there is only one nontrivial  Virasoro algebras with generators:
\bea
L_{m}=\int dx^{-} e^{2imx^{-}} T_{--}
\eea
We see that not only left and right modes decouple from 
each other, but we only have a single chiral sector.

\subsection{Quantization of the String Modes}

The general expansion for $X^{\mu}(x^{-},x^{+})$ can be written as:
\beq
X^{\mu}(x^{-},x^{+})=X^{\mu}(0)+p^{\mu}_{L}x^{-}+
p^{\mu}_{R}x^{+}+i(4\pi\alpha')^{1/2} 
\left[ \sum_{n \neq 0}\frac{\alpha_{n}^{\mu}}{n} 
e^{inx^{-}}+\int\frac{\beta_{n}^{\mu}}{n} e^{inx^{+}}dn \right]
\eeq
where due to  a Dirichlet boundary condition for $\mu = -, i$
there is no  $x^{+}$ dependence for $X^{-}$ and $X^{i}$, i.e.
 all $\beta_{n}^{-,i} =0$.   At the same
time there are no  any boundary conditions for $X^{+}$ and because of
this one can forget about $\beta_{n}^{+}$ oscillators.
Quantization of the string modes is done through 
the well known equal time commutation relations between density 
momentum and the spacetime coordinates
\beq
\left[\Pi^{\mu}(x^{-},x^{+}), X^{\nu}(x'^{-},x^{+}) \right]=
i \eta^{\mu \nu} \delta(x^{-}-x'^{-}) 
\eeq
where  $x^{+}$ is  the  world-sheet time coordinate. Simple
calculations give
\beq
\Pi^{\mu}=\frac{1}{4\pi\alpha'}\partial_{-} X^{\mu}=
\frac{1}{4\pi\alpha'} \left[ p^{\mu}_{L}-\sum_{n \neq 0} 
\alpha_{n} e^{inx^{-}} \right]
\eeq
The non-vanishing canonical commutation relations follows straightforwardly
\bea
\left[ p^{\mu}_{L},X^{\nu}(0) \right] =i \eta^{\mu \nu}\nonumber\\
\left[ \alpha_{n}^{\mu},\alpha_{m}^{\nu} \right] =
m \eta^{\mu \nu} \delta_{m+n,0}
\eea
with all the other vanishing, 
including $\beta^{+}_{n}$ with itself
\beq
\left[ \beta_{+}^{\mu},\beta_{+}^{\nu} \right]=0
\eeq
Let us make usual identifications
\beq
a_{n}^{\mu \dag}=\frac{1}{\sqrt n} 
\alpha_{-n}^{\mu}\ ~~~~~\ a_{n}^{\mu}=\frac{1}{\sqrt n} 
\alpha_{n}^{\mu}\ ~~~~\ for\ n > 0
\eeq
and  notice that we  have only left-moving 
quantized oscillators.
 We thus only have the quantized left sector of the 
Virasoro algebra. 

\section{Chiral closed strings from open strings:  
Lorenz versus Euclidean picture}

In this section, we will make a correspondence between open strings
and chiral closed strings. That is done by making a Wick rotation on
 the world-sheet as well as in target-space. 
 As we saw 
the coordinate $x^{+}$ is  a light-cone time coordinate 
and one could try to make a  Wick rotation along  $x^{+}$. 
This leaves $x^{-}$ unchanged. However this  is not a consistent 
continuation. Let us look at the world-sheet metric
\beq
ds^{2}=dx^{-}dx^{+}
\eeq
where the boundaries are $x^{-}=0$ and $x^{-}=\pi$. We cannot  get
real metric by  rotating only $x^{+}$, so the only way  is continue a 
world-sheet time coordinate $\tau \rightarrow i\tau$ and not on the
light-like time $x^{+}$. Because $\tau = x^+ - x^-$ and $\sigma = 
 x^+ + x^-$,  we  immediately get the following relation between 
 $x^{-}$ and $x^{+}$
\beq
x^{-}=x^{+ \ast}
\eeq
 which means that  we get ordinary complex coordinates 
\bea
x^{-}=z\nonumber\\ 
x^{+}=\bar{z}
\eea
An interesting  fact is 
 that the previous light-like boundaries of a world-sheet with
 Lorenz signature are now mapped to points on a  complex plan:
\bea
x^{-}=0 \rightarrow z=\bar{z}=0\nonumber\\
x^{-}=\pi \rightarrow z=\bar{z}=\pi
\eea
We end up with no boundaries on complex worksheet and so our {\it open
strings}  are mapped into 
{\it chiral closed strings}. 
With a Euclidean signature world-sheet, our action is written in the complex form
\beq
S_{P}+ S_{B}=
\frac{1}{2\pi\alpha'}\int_{\Sigma}dzd\bar{z}G_{\mu \nu}\partial_{z}
X^{\mu} \partial_{\bar{z}}X^{\nu} + V(0) - V(\pi)
\eeq
where the boundary term disappeared due to  the simple fact that boundaries
  shrinked into points. Instead  of boundary terms  we have  vertex operators
located at these  points.The operator at point $a$ is given by
\beq
V(a)=\frac{i}{8\pi\alpha'}\oint dz 
\left( X^{*}\partial_{z}X - X\partial_{z}X^{*} \right)
\eeq
where the contour integral is taken over an infinetezimal contour
 around $a$.
Since our variables are complex, 
the fields $X^{\pm}$ must also be complex and so we need to carry a
Wick rotation also in
 target space-time $T \rightarrow iT $ 
and one finds that space-time light-cone coordinates 
are complex conjugate to each other
\beq
X = X^{-}=R-iT=(R+iT)^{\ast}=(X^{+})^{\ast}
\eeq
As we see, all the unquantized oscillators $\beta^{+}_{n}$ and
momentum $p^{+}$ of the 
light-cone coordinate $X^{+}$ vanishes after the spacetime Wick
rotation. Since world-sheet boundaries have also disappear, we arrive
to a closed string configuration with only left moving oscillator
modes. Strings frozen to H-branes are thus mapped by a Wick rotation
to closed chiral strings as we only have one sector of the Virasoro
algebra in both cases.

Let us briefly discuss what happens when we have non-trivial
spacetime metric in our model. The action takes the form
\beq
S_{P+B}=-\frac{1}{4\pi\alpha'}\int_{\Sigma} dx^{-} dx^{+}
 G_{\alpha
\beta}\gamma^{ab}\partial_{a}X^{\alpha}\partial_{b}X^{\beta}-\frac{1}{4\pi\alpha'}
\int_{\partial \Sigma}dx^{+} G_{+-}X^{+}\partial_{x^{+}}X^{-}
\eeq
and the world-sheet metric will be $ds^{2}=dx^{-}dx^{+}$. 
It is easy to write  equations of motion
\beq
\partial_{x^{-}}\partial_{x^{+}}X^{\mu}+
\Gamma^{\mu}_{\alpha \beta}\partial_{x^{-}}X^{\alpha}\partial_{x^{+}}X^{\beta}=0
\eeq
and the boundary conditions
\bea
&&\left[2G_{ij}\partial_{x^{+}}X^{j}+G_{+-},_{i}X^{+}
\partial_{x^{+}}X^{-}\right]\delta X^{i}=0\nonumber\\
&&\left[G_{+-}\partial_{x^{+}}X^{-}+G_{+-},_{+}X^{+}
\partial_{x^{+}}X^{-}\right]\delta X^{+}=0\nonumber\\
&&\left[G_{+-},_{+}X^{+}\partial_{x^{+}}X^{+}+
G_{+-},_{i}X^{+}\partial_{x^{+}}X^{i}\right]\delta X^{-}=0
\eea
The indices (i,j) denote the transverse spacetime coordinates where
now explicit write partial 
derivatives with respect to the world-sheet coordinates (denoted by
$\partial_{x^{\pm}}$) 
and to the spacetime light-like coordinates (denoted by $,_{\pm}$). 
We now impose Dirichlet boundary condition for the light-cone
coordinate $X^{-}$ 
as well as for the transverse coordinates $X^{i}$. 
It is important to note that the chiral solutions 
for both $X^{-}$ and $X^{i}$ coordinates are still valid in a case
of curved space-time. 
Even if is not the most general solution, the consequence of such a 
particular solution is that the relation between 
Dirichlet boundary condition on world-sheet and the presence of
H-brane 
in curved space-time is hold:
\bea
&&\partial_{+}X^{-}\mid_{\partial \Sigma}=0 \Rightarrow 
\delta X^{-}\mid_{\partial \Sigma}=0\nonumber\\
&&\partial_{+}X^{i}\mid_{\partial \Sigma}=0 \Rightarrow 
\delta X^{i}\mid_{\partial \Sigma}=0
\eea
 Moreover, we have no boundary conditions for the $X^{+}$ light-cone
coordinate and so open strings frozen on H-brane can oscillate outside
the brane which can be  interpreted as quantum fluctuations of
H-branes.

Let us note that this picture has a direct relation to the mass spectrum of
a Schwarzshild black hole. 
After Euclidean continuation, open strings attached to H-branes become
chiral closed strings that wound around the horizon (see Fig.5), 
where now the complex time coordinate is to be identified as an
angular variable with periodicity of
 the inverse temperature $\beta$ of the black hole. 
In the case of a Schwarzshild black hole with mass $M$ the identification is,
\beq
iX^{0} \sim iX^{0} + \beta
\eeq
where in units of Planck mass $\beta$ is related to the horizon radius 
\beq
\beta=2 \pi R_{+}=\frac{4 \pi M}{M_{P}^{2}}
\eeq
 It was shown in \cite{kogan1} (see also \cite{kogan2}) that  chiral
sector  can  exist only
for quantized black holes with the discrete spectrum  $R_{+}^{2}=n$, 
i.e, the Schwarzshild black hole mass spectrum $M=M_{P}\sqrt n$.
 which is precisely the   spectrum  we have discussed in the
introduction.
\begin{figure}
\hspace{160mm}
\epsfxsize=160mm
\epsffile{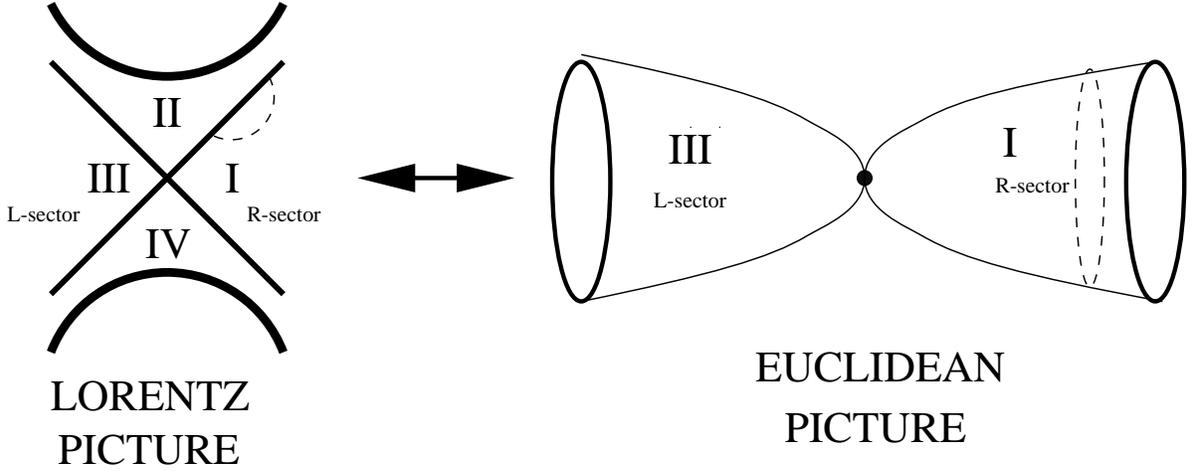}
\caption{Open strings attached to H-branes 
in Lorenz picture becomes chiral closed string in Euclidean picture}
\label{figure:Fig5}
\end{figure}


\section{$X^{-}$, $X^{+}$ as a Phase Space: non-commutative geometry
from H-branes without $B_{\mu \nu}$ field and squeezed
boundary states.}

In this section, we shall demonstrate that   $X^{+}$ and  $X^{-}$ 
 light-cone coordinates can be treated as a canonical pair in a phase
space.  To see it we shall study the dynamics  of  open string
endpoints frozen  along   H-brane in  the $X^{+}$ - direction. 
 We shall follow an analogy between open strings with  
 extra boundary action and a dissipative quantum mechanics (DQM)  
 \cite{cl}
discussed  in \cite{callan1,callan2,callan3,callan4}. 

By the very nature of the extra boundary term in curved spacetime, 
\beq
S_{B}=\frac{1}{8\pi\alpha'}\int_{\partial \Sigma} dx^{+}\ G_{-+} 
\left( X^{+} \partial_{x^{+}}X^{-}- X^{-} \partial_{x^{+}}X^{+} \right)
\eeq
it resemble to an action describing a massless charged particle in a
constant magnetic field. 
Such term is familiar in string theory where the boundary action is
given by the Wilson 
line and non-commutative geometry can arise in configurations where
the 
NS-NS $B_{\mu \nu}$ field is large (see for example \cite{seiberg}
 and references therein). 
In our situation, the role played by a background  magnetic field 
is played  by the $+-$ components of the spacetime metric $G_{-+}$.
In this paper we only discuss flat space-time (like Rindler), the 
  general case will be discussed in a future publication.

Let us again introduce parameter $\beta$ in our action 
\beq
S_{\beta}=S_{P}+\beta S_{B}
\eeq
If we start from $\beta=0$, 
we have the Polyakov action describing free strings in Minkowski
space-time with the world-sheet as the usual one with time-like
boundaries. Ordinary  open strings carry both left and right oscillating modes 
 and any boundary state $|B\rangle$ must satisfy the condition 
 which relates  Virasoro generators from both sectors 
\beq
(L_{n}-{\bar L}_{-n})|B\rangle = 0
\label{oldLn}
\eeq
This condition  means that the 
 boundary does not destroy conformal symmetry and
there is no energy flux through the boundary \cite{cardy}. 
 For a boundary state associated with  the Neumann boundary condition we have\cite{callan1},
\beq
|B\rangle_{N} = exp \left(-\sum_{m=1}^{+ \infty}\frac{1}{m}\alpha_{-m}
\cdot\tilde{\alpha}_{-m} \right)|0 \rangle
\eeq
and for a state associated to a Dirichlet boundary condition we have
\beq
|B\rangle_{D} = exp \left(+\sum_{m=1}^{+ \infty}\frac{1}{m}\alpha_{-m}
\cdot\tilde{\alpha}_{-m} \right)|0 \rangle
\eeq
All these 
 states belongs to a
 Hilbert space ${\cal H_{L}}\otimes{\cal H_{R}}$ where the vacuum
carries 
both left and right vacuum sectors,
\beq
|0\rangle = |0\rangle_{L} \otimes |0\rangle_{R}
\eeq
Let us  consider the Neumann state, for example.
We see that by increasing $\beta$ 
the initial Neumann boundary conditions for light-cone coordinates
 are replaced by
\bea
\partial_{\sigma}X^{+}+\beta\partial_{\tau}X^{+}=0\nonumber\\
\partial_{\sigma}X^{-}-\beta\partial_{\tau}X^{-}=0
\eea
We are still using world-sheet metric $ds^{2}=-d\tau^{2}+d\sigma^{2}$ 
with time-like boundaries.
By introducing the parameter $\beta$ 
we arrive at a coherent string state that is
 analogous to a path integral in the 
DQM system in a magnetic field \cite{callan2,callan3,callan4} where 
the parameter $\beta$ is proportional to the magnetic field. 
The two point function was calculated in  \cite{callan3,callan4} and
 in  the  case of  real valued light-cone target-space coordinates
 $X^{\pm}$ one gets in  the limit $\beta \rightarrow 1$
\beq
\langle 0| X^{+}(\tau, \sigma) X^{-}(\tau', \sigma) |0 \rangle = 
-\frac{a}{a^{2}+b^{2}}log(\tau-\tau')^{2} 
\delta_{-+}-\frac{\pi b}{a^{2}+b^{2}}sign(\tau-\tau')\epsilon_{-+}
\eeq
where $a^{-1}=2\pi\alpha'$ and $b^{-1}=8\pi^{2}\alpha'$. 

The first term on right hand side measures the delocalization of the
string shape. In condensed matter physics literature
  the quantity that measures a delocalization  is  called mobility. 
The logarithmic growth is a transition between two extreme limits: 
long-time behaviour is bound by a constant or it grows without limit. 
The coefficient in front  of the logarithm is the value of the critical mobility.

The second term can be interpreted as a Hall effect for
our target space-time fields. 
However, in a conformal field theory, one can compute commutators of
operators from the short 
distance behaviour of T-product (famous BJL relation, for recent
discussion see for example \cite{schomerus},\cite{seiberg} and references therein)
 and  after simple calculations one gets
\beq
[X^{+}(\tau),X^{-}(\tau')]=T 
\left( X^{+}(\tau)X^{-}(\tau^{-})-X^{+}(\tau)X^{-}(\tau^{+}) \right)=
\Theta \epsilon_{-+}
\eeq
that is, the boundary term we introduced to construct $H$-brane at the 
same time  tells us  that  $X^{+},X^{-}$ can be seen as coordinates
 in  a non-commutative spacetime, with non-commutativity parameter
$\Theta=\frac{\pi b}{a^{2}+b^{2}}$.

When $\beta=1$ we get H-brane. Using the same method as in
\cite{callan1}  we can construct a boundary state  associated with
H-brane where instead  of (ref{oldLn})  we have chiral constraint  
$ L_{n}|H\rangle = 0 $

To find $|H\rangle$ we shall use the fact  that $\partial_{+}X^{\mu} = 0$
for $\mu=-,i$.
and 
\beq
X^{\mu}(x^{-},x^{+}) = X^{\mu}(0) + p_{L}^{\mu}x^{-} +
i(4\pi\alpha')^{1/2}\sum_{n \neq 0}\frac{\alpha_{n}}{n}e^{inx^{-}}
=  X^{\mu}(0) + i(4\pi\alpha')^{1/2}\sum_{n > 0} \frac{1}{\sqrt{n}}X^{\mu}_{n}
\eeq
where 
\beq
X^{\mu}_{n}=a_{n}^{\mu}e^{inx^{-}}-a_{n}^{\mu \dag}e^{-inx^{-}}
\eeq
Let us consider states
\beq
|X\rangle = exp \left(
-\frac{1}{4}(X|X)-\frac{1}{2}(a^{\dag}|a^{\dag})-(X|a^{\dag}) \right)
 |0\rangle_{L}
\eeq
where we use  the same notation as \cite{callan1}
\beq
(X|X)=\sum_{\mu = 0}^{25}\sum_{m=1}^{\infty}X_{m}^{\mu}X_{m}^{\mu}
\eeq
These  states satisfy the eigenvalue condition
\beq
(-a_{n}^{\dag}+a_{n}-X_{n}) | X \rangle = 0
\eeq
as well as the completeness relation
\beq
\int_{-\infty}^{+\infty} {\cal D}X |X \rangle \langle X|=1
\eeq
Taking $H$-brane  at $x^{-}=0$ the  boundary state associated with it
is given by 
\bea
&&|H \rangle = \int {\cal D}X |X\rangle\nonumber\\
&&~~~=\int {\cal D}X exp \left(
-(\frac{1}{2}X+a^{\dag}|\frac{1}{2}X+a^{\dag}) +
\frac{1}{2}(a^{\dag}|a^{\dag}) \right) |0 \rangle_{L}
\eea
and  after simple  gaussian integration one gets the boundary state
\beq
|H\rangle = exp \left( \frac{1}{2}\sum_{m=1}^{+\infty} 
\frac{1}{m}\alpha_{-m}\cdot \alpha_{-m} \right)|0\rangle_{L}
\eeq
where $\alpha_{-n} = \sqrt n a_{n}^{\dag}$
We stress that in spite of similarity between the previous three
states, $|B\rangle_{N}$, $|B\rangle_{D}$ and $|H\rangle$ 
we see that this last state is:
\begin{itemize}

\item{ a chiral state}

\item { a squeezed state }

\end{itemize}

Those two properties are to be compare with the analogous properties
of the first two boundary states, that are non-chiral and coherent states. 
 Squeezed quantum states known for a long time 
 in quantum optics and measurement theory ( for a review
 of squeezed states see, for example \cite{sqrev}). 
 The simplest one-mode squeezed state is parameterized by the
 two parameters $r$ and $\phi$ and can be obtained by  acting
 on the vacuum  with the unitary squeezing operator $S(\xi)$
\begin{equation}
|\xi> = S(\xi)|0> =
 \exp\left[\frac{1}{2}\left(\bar{\xi}a^{2} - \xi (a^{\dagger})^2)
\right)\right]
\end{equation}
where $\xi = r \exp(i\phi)$ is the squeezing parameter.
The mean number of quanta in  the squeezed state is
$\bar{N}= \sinh ^{2} r$.
To see this let us note that using the squeezing operator $S$ one
 can make the Bogolyubov transformation
$b = SaS^{\dagger},~~~
b^{\dagger} = Sa^{\dagger}S^{\dagger}$
 and after some algebra one gets:
\begin{eqnarray}
b &  =  & \cosh r a + \exp(i\phi) \sinh r a^{\dagger},
~~~~~~~
b^{\dagger}   = \exp(-i\phi) \sinh r a + \cosh r a^{\dagger}
 \nonumber \\
a &  =  & \cosh r b - \exp(i\phi) \sinh r b^{\dagger},
~~~~~~~
a^{\dagger}  = -\exp(-i\phi) \sinh r a + \cosh r a^{\dagger}
\label{abrelation}
\end{eqnarray}
The new operator $b$ is the annihilation operator for the
 squeezed state
\begin{eqnarray}
b|\xi> = b S |0> = S a S^{\dagger} S |0> = S a |0> = 0
\end{eqnarray}
 Then it is easy to see that
\begin{eqnarray}
\bar{N} = <\xi|a^{\dagger}a|\xi>
= \sinh^{2} r <\xi|bb^{+}|\xi> = \sinh^{2} r
\end{eqnarray}
After some algebra (see for example \cite{amadokogan} and references
therein)  one can write
the normalized squeezed state as
\begin{equation}
|\alpha> = (1-|\alpha|^2)^{1/4} \exp \left( \frac{\alpha}{2}
 (a^{\dagger})^2 \right) |0>
\end{equation}
In ours case  $\alpha= 1$ and our state can not be normalized,
 but it is interesting that it is just at the border between
normalizable and non-normalizable states.
A squeezed state is a minimum uncertainty state as well as a coherent 
state. Contrary to coherent states which  have  minimal quantum
uncertainty for both conjugate variables
 for  the squeezed state one can
 get  large  uncertainty for one variable 
while the other is ``squeezed" to keep the product fixed.
 It seems that this interesting property is related to the very nature 
of localization of open strings on H-brane. It is also interesting
that in this state we have only pairs of open string oscillators.
 More detailed analysis of this state will be given in separate publication.

 Let us note that one can get a density matrix starting from dynamics
in both sectors I and III and then mapping the  wave function
\beq
|\Psi\rangle = \sum_m A_m |m\rangle_{L} \otimes |m\rangle_{R}
\eeq
into density matrix
\beq
\rho = \sum_m A_m  |m\rangle_{R}\langle m |_{R}
\eeq
as been suggested in \cite{thooft84} 
\footnote{ when this paper was prepared for
publication   we became aware about recent paper 
\cite{hepth/0106112}  in which density matrix from wave
function on maximally   extended eternal  AdS black hole was discussed}  
It is quite interesting that  besides squeezed state we have discussed 
 it may be another candidate for a quantum state of H-brane, but in
this case it will have non-zero entropy
\beq
S = - Tr \rho\ln\rho = -\sum_m A_m \ln A_m
\eeq
We hope to return to this issue in the future publication and see if
the entropy of H-brane in a mixed state can explain entropy of quantum 
horizon.


\section{Conclusion}

In  this paper we  suggested
 a simple model where open string endpoints are fixed in a single null
direction. We conjectured  that this gives us a  new class of branes -
 $H$-branes. The excitations of H-branes are chiral open strings.
 These strings are  not just  Schild null-strings \cite{schild}, 
where all points of a string (and not only it's endpoints) travel at
the speed of light. 
The fact that  H-branes are associated with a  single  chiral sector on
world-sheet gives us a totally new  boundary conditions. It is
interesting to analyze in more details relations between these
conditions  and  usual boundary
conditions for boosted D-branes \cite{cardy,polchinski1} and also to
 recently introduced  A-branes and B-branes \cite{schomerus2,maldacena3}.

We conjecture  that H-branes play important role in stringy
description  of quantum horizons, such as black hole or cosmological
horizons. $H$-branes has a remarkable property that after analytical
continuation  one gets chiral closed strings in near-horizon Euclidean 
geometry, which have been introduced some time ago
\cite{kogan1,kogan2}. At the same time  they may give us 
 a space/phase space transmutation and  
 non-commutativity of light-cone coordinates.
It will be quite natural to conjecture that $H$-branes ultimately
related to Bekenstein-Hawking entropy. These and related questions
will be discussed in future publications.


\section{Acknowledgments}
One of us (IIK) would like to thank T.Damour about interesting  and
stimulating  discussions about strings and black holes. 
The work of IIK  is
supported in part by the PPARC rolling grant PPA/G/O/1998/00567, by
the EC TMR grants  HRRN-CT-2000-00148 and  HPRN-CT-2000-00152.
We would like to thank S. Kawai, B. Tekin and M. Costa for useful
discussions. 
The work of NBR is supported by Praxis XXI/BD/18138/98 grant from FCT (Portugal).


\end{document}